\documentclass[aps,prl,superscriptaddress, twocolumn, amsmath,amssymb,floatfix,longbibliography]{revtex4-1}
\usepackage{graphicx}

\usepackage{bm}
\usepackage[colorlinks=True, allcolors=blue]{hyperref}
\usepackage{siunitx}
\usepackage{balance}
\usepackage{color,soul}

\begin{document}
    \title{Tuning the glass-forming ability of metallic glasses through energetic frustration}

    \author{Yuan-Chao Hu}
    \affiliation{Department of Mechanical Engineering \& Materials Science, Yale University, New Haven, Connecticut 06520, USA}

    \author{Jan Schroers}
    \affiliation{Department of Mechanical Engineering \& Materials Science, Yale University, New Haven, Connecticut 06520, USA}

    \author{Mark D. Shattuck}
    \affiliation{Benjamin Levich Institute and Physics Department, The City College of New York, New York, New York 10031, USA.}

    \author{Corey S. O'Hern}
    \email[]{corey.ohern@yale.edu}
    \affiliation{Department of Mechanical Engineering \& Materials Science, Yale University, New Haven, Connecticut 06520, USA}
    \affiliation{Department of Physics, Yale University, New Haven, Connecticut 06520, USA.}
    \affiliation{Department of Applied Physics, Yale University, New Haven, Connecticut 06520, USA.}
    \date{\today}

\begin{abstract}
The design of multi-functional BMGs is limited by the lack of a
quantitative understanding of the variables that control the
glass-forming ability (GFA) of alloys.  Both geometric frustration
(e.g. differences in atomic radii) and energetic frustration
(e.g. differences in the cohesive energies of the atomic species)
contribute to the GFA. We perform molecular dynamics simulations of
binary Lennard-Jones mixtures with only energetic frustration. We show
that there is little correlation between the heat of mixing and
critical cooling rate $R_c$, below which the system crystallizes,
except that $\Delta H_{\rm mix} < 0$.  By removing the effects of
geometric frustration, we show strong correlations between $R_c$ and
the variables $\epsilon_- =
(\epsilon_{BB}-\epsilon_{AA})/(\epsilon_{AA}+\epsilon_{BB})$ and
${\overline \epsilon}_{AB} =
2\epsilon_{AB}/(\epsilon_{AA}+\epsilon_{BB})$, where $\epsilon_{AA}$
and $\epsilon_{BB}$ are the cohesive energies of atoms $A$ and $B$ and
$\epsilon_{AB}$ is the pair interaction between $A$ and $B$ atoms.  We
identify a particular $f_B$-dependent combination of $\epsilon_-$ and
${\overline \epsilon}_{AB}$ that collapses the data for $R_c$ over
nearly $4$ orders of magnitude in cooling rate.
\end{abstract}



    \maketitle

Bulk metallic glasses (BMGs) are amorphous alloys that possess
promising structural, mechanical, and functional
properties~\cite{JanPT,Demetriou2011,WANG2012487}. However, a given
BMG may not possess multiple desirable properties, such as high
elastic strength and biocompatibility in the case of BMGs
used in biomedical applications~\cite{Zberg2009}. Thus, {\it de novo}
design of BMGs with multi-functional properties is an important
goal. A key impediment to progress is that one cannot currently
predict the glass-forming ability (GFA) of a given
alloy~\cite{LU20023501}.  The most prominent and widely used features
for identifying BMGs were suggested by Inoue in
2000~\cite{INOUE2000279}: 1) BMGs are typically multicomponent systems
consisting of three or more elements, 2) the size ratios of the three
main constituents differ by more than $12\%$, and 3) the heat of
mixing $\Delta H_{\rm mix}$ among the three main elements is
negative. However, there are many examples of metallic
glasses that do not obey these rules. First, several binary
alloys (such as CuZr) possess GFAs that are comparable to those for
multi-component BMGs~\cite{XU20042621,Li1816,mei2004binary}. Also,
there are many ternary alloys (e.g. Al, Cu, and V) that have 
$R_c < 10^6$ K/s, but the diameter ratios among
the three elements differ by less than $12\%$~\cite{tsai1988ductile}.
Further, recent experimental studies have shown that even monoatomic
metallic systems can form glasses via rapid
cooling~\cite{Zhong2014}. Thus, it is clear that a more quantitative
theoretical framework is necessary for predicting the GFA of alloys.

There are two main contributions to the GFA of an alloy, geometric and
energetic frustration~\cite{KZ2013,Tanaka2006NP}.  Geometric
frustration can be achieved in alloys using elements with sufficiently
different metallic radii~\cite{KZ2013,Miracle2004, Sheng2006}, which
allows the glass phase to pack more desely, but severely strains the
competing crystalline phases. Energetic frustration can be achieved in
alloys even with elements of similar sizes, if they possess different
cohesive energies and strong interactions between different atomic
species. While there have been many computational studies of the
variation of $R_c$ with geometric
frustration~\cite{HU2015NC,Tanaka2006NP,ChengPRL2009}, there are few
studies that have investigated how energetic frustration in the
absence of geometric frustration affects the GFA.

In this Letter, we carry out molecular dynamics simulations of binary
Lennard-Jones (LJ) mixtures with atoms of the same size, but different
cohesive energies, to understand the critical cooling rate as a
function of the degree of energetic frustration. We find several
important results: 1) We show that there is little correlation between
the GFA and heat of mixing in binary and multi-component metallic
glass formers.  2) Instead, we find that there is a particular
combination of the difference in the cohesive energies and the pair
interactions among different species in binary alloys that yields the
best GFA for each composition. 3) We rationalize these findings for
binary LJ systems with the best GFA by considering separation
fluctuations and chemical ordering~\cite{chemicalorder} among nearest
neighbor atoms.

We focus on binary LJ mixtures in three dimensions
with vanishing geometric, but tunable energetic frustration. The pairwise
interaction potential is:
\begin{equation}
\label{lj}
V(r_{ij}) = 4\epsilon_{ij} \left[ \left( \frac{\sigma}{r_{ij}}\right)^{12} - \left( \frac{\sigma}{r_{ij}}\right)^6 \right], 
\end{equation}
where $\sigma$ is the diameter of atoms $A$
and $B$, $r_{ij}$ is the separation between atoms $i$ and $j$,
$\epsilon_{AA}$ and $\epsilon_{BB}$ are the cohesive energies of atoms
$A$ and $B$, and $\epsilon_{AB}$ is the interaction energy between $A$
and $B$. The potential is truncated and shifted at $r_{ij} =
2.5\sigma$, and the total potential energy is $V=\sum_{i>j}
V(r_{ij})$. We consider $N = N_A + N_B = 2000$ atoms with equal mass
$m_A=m_B$ in a cubic box and periodic boundary
conditions in all directions. Length, energy, pressure, and time
scales will be reported in units of $\sigma$, $\epsilon_{AA}$,
$\epsilon/\sigma_{AA}^3$, and $\sqrt{m_A \sigma/\epsilon_{AA}}$.

We first equilibrate each system with a fraction of $B$ atoms, $f_B =
N_B/N$, and combinations of $\epsilon_{BB}/\epsilon_{AA}$ and
$\epsilon_{AB}/\epsilon_{AA}$ at high temperature $T=5.0$ (using a
Nose-Hoover thermostat~\cite{nose,hoover}) and then quench them to low
temperature $T=0.01$ as a function of cooling rate $R$. The thermal
quenches are performed at fixed pressure $P_0=10$ to avoid
cavitation~\cite{npt}. We find that the particular value of $P_0$ does
not strongly affect the GFA in systems that do not cavitate over the
range $10^{-2} < P_0 < 10$. (See Supplemental Material.)

\begin{figure}[h]
  \includegraphics[width = 8.5 cm]{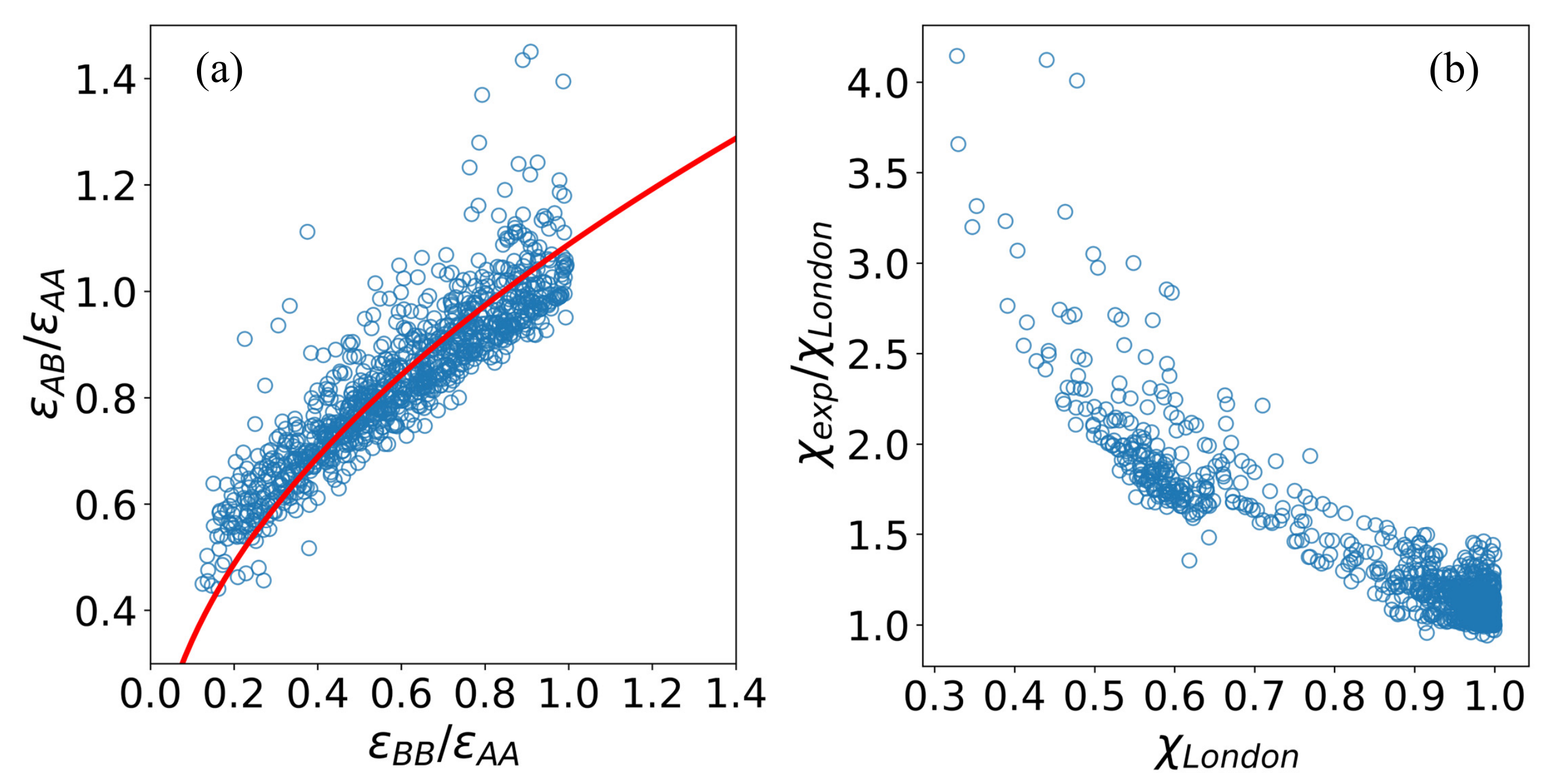}
\caption{(a) The interaction energy $\epsilon_{AB}$ (normalized by $\epsilon_{AA}$) from the
pairwise heat of mixing $\Delta H_{p}$ plotted versus the cohesive
energy ratio $\epsilon_{BB}/\epsilon_{AA}$ for $990$ binary alloys
involving $45$ elements found in metallic glasses~\cite{cohesive,AA2005}. We
chose element $A$, such that $\epsilon_{AA} > \epsilon_{BB}$. The solid 
line obeys $\epsilon_{AB} =
c \sqrt{\epsilon_{BB} \epsilon_{AA}}$ with $c=1.09$.  (b) The ratio
of $\chi_{\rm exp} = \epsilon_{AB}/\sqrt{\epsilon_{AA}
\epsilon_{BB}}$ to the London expression, $\chi_{\rm London}$ in 
Eq.~\ref{London}, plotted versus $\chi_{\rm London}$ for the same data
in (a).}
\label{fig1}
\end{figure}

To understand the relevant range of parameter space for the cohesive
energies, $\epsilon_{AA}$ and $\epsilon_{BB}$, and interaction energy
$\epsilon_{AB}$, we cataloged these values for $990$ binary alloys
involving $45$ elements that occur in metallic glasses. For this
analysis, we chose element $A$ such that $\epsilon_{AA} >
\epsilon_{BB}$ and used the pairwise definition of the heat of mixing,
$\Delta H_p(i,j) = (\epsilon_{ii} +\epsilon_{jj})/2 - \epsilon_{ij}$,
to calculate $\epsilon_{AB}$~\cite{AAMix}. Values for $\epsilon_{AA}$,
$\epsilon_{BB}$, and $\Delta H_p$ were obtained from experimental
data~\cite{cohesive,AA2005}.  In Fig.~\ref{fig1} (a), we show that
binary alloys exist over a narrow range of parameters, $0.5
\lesssim \epsilon_{AB}/\epsilon_{AA} \lesssim 1.4$ and $0.1 \lesssim
\epsilon_{BB}/\epsilon_{AA} < 1$. In contrast, these energetic
parameters can exist over a wider range in ionic liquids and molten
salts~\cite{ionicliquid, molten}. Albeit with scatter, the
experimental data scales as $\epsilon_{AB} \propto \sqrt{\epsilon_{AA}
  \epsilon_{BB}}$, which is similar to the London mixing rule
$\epsilon_{AB} = \chi_{\rm London} \sqrt{\epsilon_{AA}
  \epsilon_{BB}}$~\cite{london}, where
\begin{equation}
\label{London}
\chi_{\rm {London}} = \frac{2 \sqrt{I_{A} I_{B}}}{I_{A}+I_{B}}
\left[ \frac{2 \sqrt{\sigma_{AA} \sigma_{BB}}}{\sigma_{AA}+\sigma_{BB}}  
\right]^6, 
\end{equation}
$\sigma_{ij} = (\sigma_i+\sigma_j)/2$ is the average diameter of atoms
$i$ and $j$, and $I_{A}$ and $I_{B}$ are the ionization energies of
atoms $A$ and $B$. In Fig.~\ref{fig1} (b), we show the
ratio of $\chi_{\rm exp} = \epsilon_{AB}/\sqrt{\epsilon_{AA}
  \epsilon_{BB}}$ for the experimental data to $\chi_{\rm
  London}$. More than $70\%$ of the data obeys the London mixing rule
with $1 < \chi_{\rm exp}/\chi_{\rm London} < 1.25$. To more
fully understand the effects of energetic frustration on the GFA of 
binary mixtures, below we
independently vary $\epsilon_{AB}/\epsilon_{AA}$ and
$\epsilon_{BB}/\epsilon_{AA}$ over a much wider range than in
Fig.~\ref{fig1} (a).

\begin{figure*}[t]
  \includegraphics[width = \textwidth]{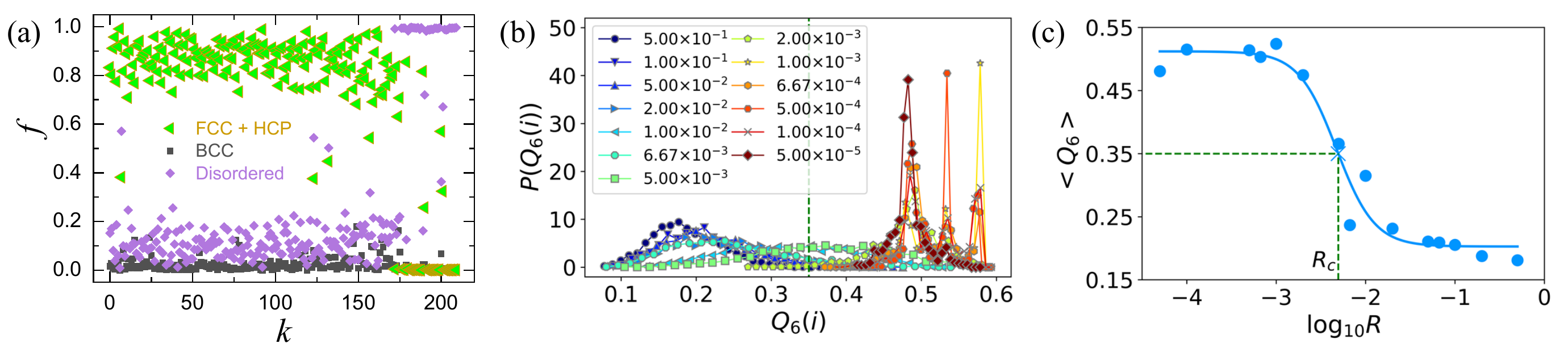}
\caption{(a) Fraction $f$ of each system (labelled $k=1,\ldots,210$) with 
a given local structure
(FCC, HCP, BCC, or disordered) for a slow cooling rate 
($R = 5 \times 10^{-5}$) in binary LJ
systems with $f_B = 0.5$ over the full range of cohesive and
interaction energies. (b) Distribution of the local bond
orientational order parameter $P(Q_6(i))$ for systems with
$\epsilon_{BB}/\epsilon_{AA}=\epsilon_{AB}/\epsilon_{AA}=1.0$ over four orders of
magnitude in cooling rate $R$.  (c) Average bond orientational 
order $\langle Q_6 \rangle$ for the system in (b) versus
$R$.  $R_c =(\langle Q_6
\rangle_0 + \langle Q_6 \rangle_{\infty})/2$ is obtained by fitting
the data to a logistic function $(\langle Q_6 \rangle-\langle
Q_6\rangle_{\infty})/(\langle Q_6 \rangle_0 - \langle Q_6
\rangle_{\infty}) = (1 -\tanh[\log_{10}(R/R_c)^{1/\kappa}])/2$,
where $\langle Q_6 \rangle_0$ and $\langle Q_6 \rangle_{\infty}$ are
the average bond orientational order in the limits of $R \rightarrow 0$ and $\infty$, and $0 <\kappa <1$ is the stretching factor. The vertical
dashed line in (b) indicates the $\langle Q_6 \rangle$ that
determines $R_c$ (vertical dashed line in (c)).}
\label{fig2}
\end{figure*}

To quantify the GFA, we analyze the positional order of the system by
measuring the bond orientational order parameter for atom
$i$~\cite{BOO, BOOVoronoi}:
\begin{equation}
\label{Q6}
Q_6(i) =\left[\frac{4\pi}{13}\sum_{m = -6}^{m = 6} \left|\frac{1}{N_i + 1}\left(q_{6m}(i) + \sum_{j = 1}^{N_i} q_{6m}(j)\right)\right|^2\right]^{1/2},
\end{equation} 
where $q_{6m}(i) = N_i^{-1} \sum_{j=1}^{N_i} (A_j^i / A_{\rm tot}^i)
Y_{6m}(\theta(\bm{r}_{ij}), \phi(\bm{r}_{ij}))$,
$Y_{6m}(\theta(\bm{r}_{ij}), \phi(\bm{r}_{ij}))$ is the spherical
harmonic of degree $6$ and order $m$, $\theta$ is the polar angle and
$\phi$ is the azimuthal angle of the vector ${\bm r}_{ij}$ from atom
$i$ to $j$, $N_i$ is the number of Voronoi neighbors of atom $i$,
$A_j^i$ is the area of the Voronoi cell face separating atoms $i$ and
$j$, and $A_{\rm tot}^i$ is the total area of all faces of the Voronoi
cell for atom $i$ ~\cite{BOOVoronoi}. 

The bond orientational order can distinguish between disordered
systems ($Q_6 \lesssim 0.3$) and systems with
crystalline order [e.g. face-centered cubic (FCC) with $Q_6 = 0.575$,
  body-centered cubic (BCC) with $Q_6 = 0.511$, and hexagonal close
  packed (HCP) $Q_6 =0.485$].  In Fig.~\ref{fig2} (a), we show the
fraction $f$ of each sample with local FCC, HCP, BCC, and disordered
structure (using adaptive common neighbor analysis~\cite{cna}) in
systems with $f_B = 0.5$ over the full range of cohesive and
interaction energies for $R=5\times 10^{-5}$.  For
more than $80\%$ of the systems, the fraction of atoms with FCC or HCP
order exceeds $0.70$, whereas very few atoms possess BCC order. (We
verify this result for other cooling rates in Supplemental
Material.) In Fig.~\ref{fig2} (b), we plot the distribution
$P(Q_6(i))$ for a system with $\epsilon_{BB}/\epsilon_{AA} =
\epsilon_{AB}/\epsilon_{AA}=1$ and several $R$.  For
$R> R_c$, the systems are disordered and $P(Q_6(i))$ has a peak near
$Q_6 \approx 0.2$. For $R<R_c$, $P(Q_6(i))$ develops
peaks near the values corresponding to FCC and HCP order. The peak
near $Q_6(i) \approx 0.535$ corresponds to regions of adjacent FCC and
HCP order, not to BCC order as shown in Supplemental Material. In
Fig.~\ref{fig2} (c), we show that $\langle Q_6 \rangle = N^{-1}
\sum_{i=1}^N Q_6(i)$ versus $R$ is similar to a logistic function, and
$R_c$ can be
determined by $R_c = (\langle Q_6 \rangle_0 + \langle Q_6
\rangle_{\infty})/2$, where $\langle Q_6 \rangle_0$ and $\langle Q_6
\rangle_{\infty}$ are the values in the limits $R\rightarrow 0$ and
$\infty$ limits.

\begin{figure}[h]
  \includegraphics[width = 8.5 cm]{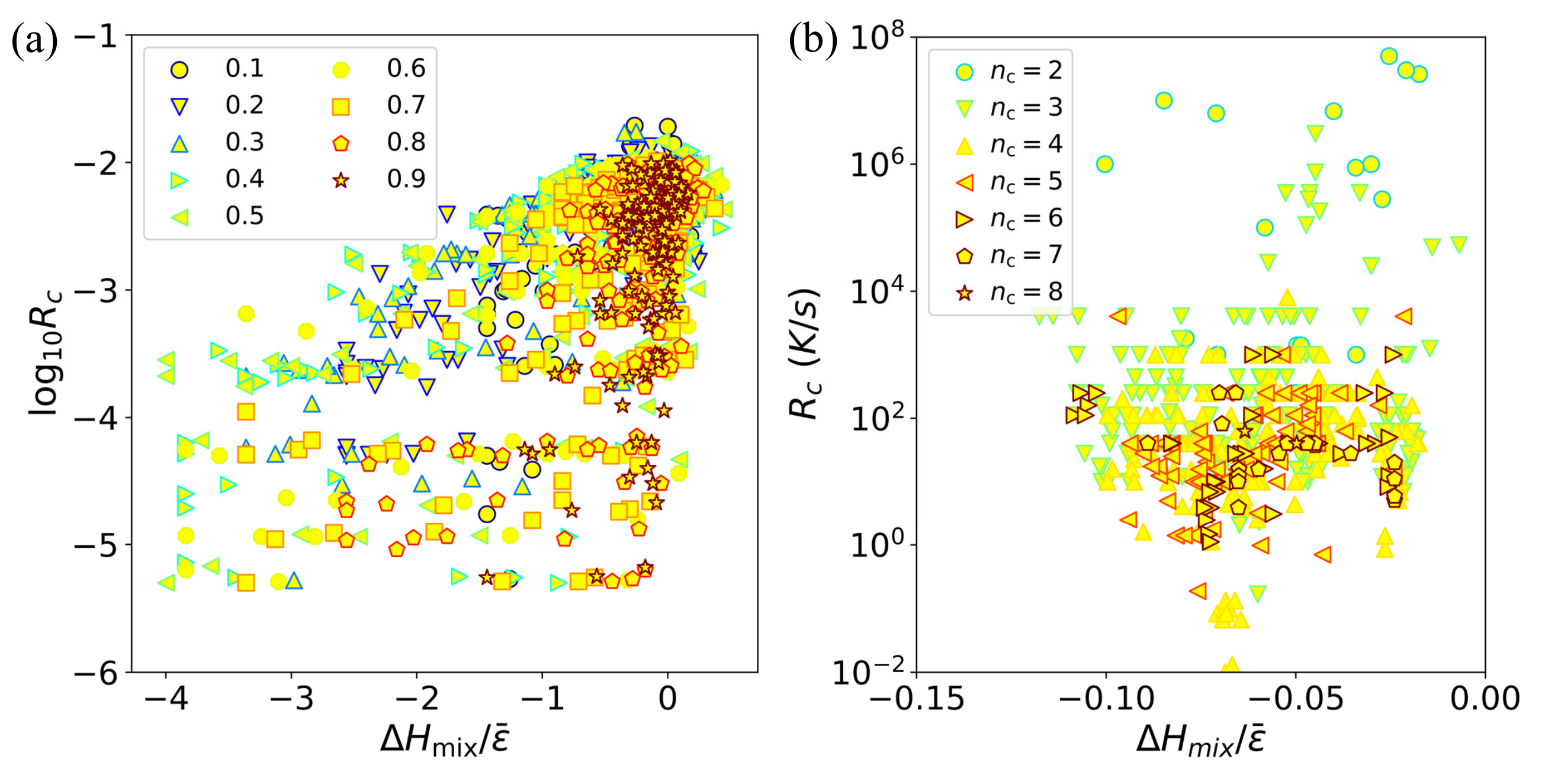}
\caption{(a) $R_c$ from simulations 
of binary LJ systems versus the heat of mixing $\Delta H_{\rm mix}/{\overline \epsilon}$, 
where ${\overline \epsilon}$ is the average cohesive energy, for nine 
values of $f_B$. (b) $R_c$ (in ${\rm K/s}$) versus $\Delta H_{\rm mix}/{\overline 
\epsilon}$ from 
experiments on $482$ 
metallic glass formers with $n=2$,$\ldots$,$8$ different atomic species.}
\label{fig3}
\end{figure}

What combination of $\epsilon_{AA}$, $\epsilon_{BB}$,
$\epsilon_{AB}$, and $f_B$ controls the GFA in alloys?  One
possibility is the heat of mixing, which can be generalized for
multi-component alloys as $\Delta H_{\rm mix} = 4 \sum_{i \ne j} f_i
f_j \Delta H_p(i,j)$~\cite{AAMix}. In Fig.~\ref{fig3} (a), we show
$R_c$ versus $\Delta H_{\rm mix}$ (normalized by the average cohesive
energy ${\overline \epsilon}$) for all binary LJ systems
studied. We find little correlation between $R_c$ and $\Delta H_{\rm
  mix}$ in the simulations ~\cite{liH}. We also assembled a database
of $482$ metallic glass formers with $n_c =2,\ldots,8$ different
atomic species (see Supplemental Material). The experimental data is
similar to the simulation data; there is no correlation between $R_c$
and $\Delta H_{\rm mix}$, other than $\Delta H_{\rm mix} < 0$ for all
metallic glasses. Note that the simulations cover a much wider range
of $\Delta H_{\rm mix}/{\overline \epsilon}$ than experiments on
metallic glasses, but $R_c$ in the
simulations corresponds to only rapid cooling, $10^{13}$ to $10^9$
${\rm K/s}$.

\begin{figure*}[t]
  \includegraphics[width = \textwidth]{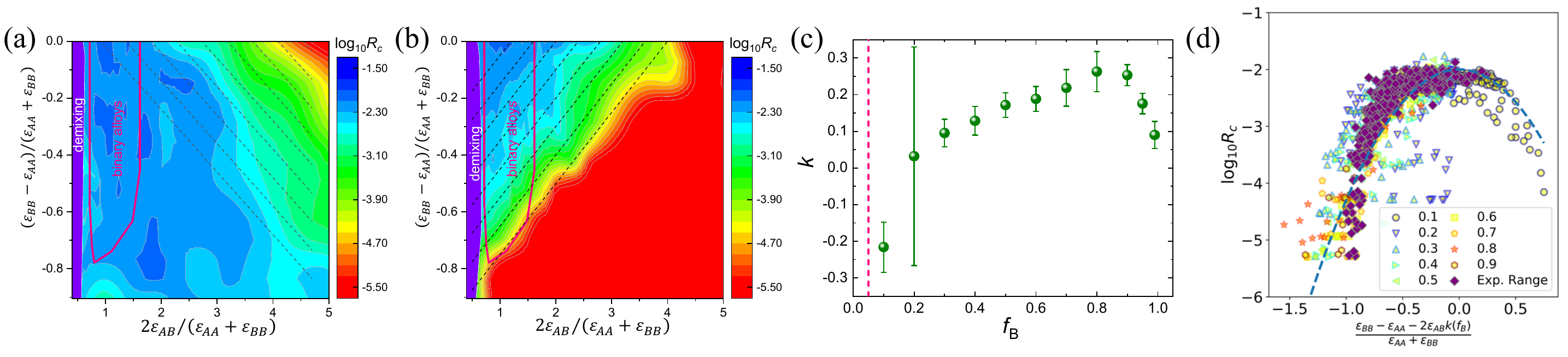}
\caption{Contour plots of equal values of $R_c$ 
(on a 
logarithmic scale decreasing from blue to red) versus 
${\overline \epsilon}_{AB} =2\epsilon_{AB}/(\epsilon_{AA} +
\epsilon_{BB})$ and $\epsilon_- =(\epsilon_{BB} -
\epsilon_{AA})/(\epsilon_{AA} + \epsilon_{BB})$ at
(a) $f_B=0.1$ and (b) $0.9$. Demixing occurs for 
$\epsilon_{AB}< (\epsilon_{AA} + \epsilon_{BB})/2$ (purple
region). The domain of experimentally accessible binary alloys is 
enclosed within the solid pink curve ({\it cf.} Fig.~\ref{fig1} (a)). 
The dashed lines represent
linear approximations of the equal-$R_c$ contours. (c) The best fit slope $k$ of the equal-$R_c$ contour lines in the 
$\epsilon_-$ and ${\overline \epsilon}_{AB}$ plane plotted 
versus $f_B$. For $f_B < 0.05$ (vertical dotted line), $R_c$ is uniform 
and $k$ is undefined. (d) $\log_{10} R_c$ 
versus $\epsilon_- -k(f_B) {\overline \epsilon}_{AB}$ for all 
systems studied. The dashed line obeys Eq.~\ref{collapse}. Binary 
LJ systems with $\epsilon_-$ and ${\overline \epsilon}_{AB}$ in the 
experimental range in Fig.~\ref{fig1} (a) are indicated by filled 
diamonds.}
\label{fig4}
\end{figure*}

In Fig.~\ref{fig4} (a) and (b), we show contour plots of $R_c$ versus
${\overline \epsilon}_{AB} = 2 \epsilon_{AB}/(\epsilon_{AA}
+\epsilon_{BB})$ and $\epsilon_-= (\epsilon_{BB} -
\epsilon_{AA})/(\epsilon_{AA}+\epsilon_{BB})$ for binary LJ systems
with $f_B=0.1$ and $0.9$.  We find strong correlations between $R_c$
and $\epsilon_-$ and ${\overline \epsilon}_{AB}$.
However, the contours of equal values of $R_c$ in the $\epsilon_-$ and
${\overline \epsilon}_{AB}$ plane are very different for $f_B = 0.1$
and $0.9$. $R_c$ increases with increasing ${\overline \epsilon}_{AB}$
and increasing $\epsilon_-$ for $f_B = 0.1$, whereas $R_c$ increases
with increasing ${\overline \epsilon}_{AB}$ and {\it decreasing}
$\epsilon_-$ for $f_B = 0.9$.  For $f_B \gg f_A$ with a majority of
$B$ atoms and only a small fraction of $A$ atoms, to have good GFA,
the cohesive interaction between $B$ atoms must be small compared to
that for $A$ atoms with $\epsilon_{BB} - \epsilon_{AA} < 0$ and the
interaction between $A$ and $B$ atoms must be strong with ${\overline
  \epsilon}_{AB} \gg 1$. Similarly, when $f_A \gg f_B$ with a majority
of $A$ atoms and only a small fraction of $B$ atoms, to have good GFA,
the cohesive interaction between $B$ atoms must be strong (or at least
comparable to that between $A$ atoms with $\epsilon_{BB} \approx
\epsilon_{AA}$) and the interaction between $A$ and $B$ atoms must be
strong with ${\overline \epsilon}_{AB} \gg 1$.  Note that the $R_c$
contours are symmetric with respect to switching the labels of atoms
$A$ and $B$, and thus we only show the region $\epsilon_{BB} -
\epsilon_{AA} \le 0$.

We approximate the $R_c$ contours as straight lines in the
$\epsilon_-$ and ${\overline \epsilon}_{AB}$ plane for each $f_B$ and
plot the slope $k$ versus $f_B$ in Fig.~\ref{fig4} (c). The slope
crosses zero near $f_B \approx 0.2$ and reaches a peak value of
$k\approx 0.25$ near $f_B \approx 0.8$.  As $f_B \rightarrow 1$, the
system becomes monoatomic with all $B$ atoms, the GFA depends only on
$\epsilon_-$, and thus $k \rightarrow 0$. As $f_B \rightarrow 0$, the
system becomes monoatomic with all $A$ atoms, and the GFA is
independent of $\epsilon_-$ and ${\overline \epsilon}_{AB}$.  In this
regime, the slope of the contours in the $\epsilon_-$ and ${\overline
  \epsilon}_{AB}$ plane is undefined as indicated by the vertical
dashed line in Fig.~\ref{fig4} (c). In
Fig.~\ref{fig4} (d), we show that the data for $R_c$ can be collapsed
by plotting $\log_{10} R_c$ versus $[\epsilon_- -k(f_B) {\overline
    \epsilon}_{AB}]$. We find that the GFA in binary LJ systems obeys
a roughly parabolic form:
\begin{equation}
\label{collapse}
\log_{10} R_c \approx c_1 [\epsilon_- -k(f_B) {\overline \epsilon}_{AB}]^2 + 
\log_{10} R_0, 
\end{equation}
where $c_1 \approx -2$ gives the concavity and $R_0 \approx 10^{-2}$
is the cooling rate in the $\epsilon_- \rightarrow 0$ and ${\overline
  \epsilon}_{AB} \rightarrow 0$ limits.

There are two striking features about the $R_c$ contours in
Fig.~\ref{fig4} (a) and (b). First, $R_c$ increases with increasing
$\epsilon_-$ and ${\overline \epsilon}_{AB}$ for small $f_B$,
indicating that systems with the best GFA possess $\epsilon_{BB} \sim
\epsilon_{AA}$ and ${\overline \epsilon}_{AB} \gg 1$. To frustrate
crystallization for small $f_B$, $\epsilon_{BB}/\epsilon_{AA}$ should
be as large as possible, approaching $\epsilon_{BB}/\epsilon_{AA}
\rightarrow 1$.  Similarly, large ${\overline \epsilon}_{AB}$ allows
the $B$ atoms to act as low mobility defects with root-mean-square (rms)
fluctuations $\Delta r_{AB} = \langle r^2_{AB} \rangle -\langle r_{AB}
\rangle^2 < \Delta r_{AA}$ in the low-temperature glass, where
$\langle r_{AB} \rangle$ is the average separation between an $A$ atom
and a Voronoi-neighbor $B$ atom.  (See the Supplemental Material.)
Second, $R_c$ increases with decreasing $\epsilon_-$ and increasing
${\overline \epsilon}_{AB}$ for large $f_B$. In this case,
$\epsilon_{BB} \rightarrow 0$ prevents $B$ atoms from
clustering. Also, in the large ${\overline \epsilon}_{AB}$ limit, the
$A$ atoms act as low mobility defects with rms
fluctuations $\Delta r_{AB} < \Delta r_{BB}$ in the low-temperature
glass.

\begin{figure*}[t]
  \includegraphics[width = \textwidth]{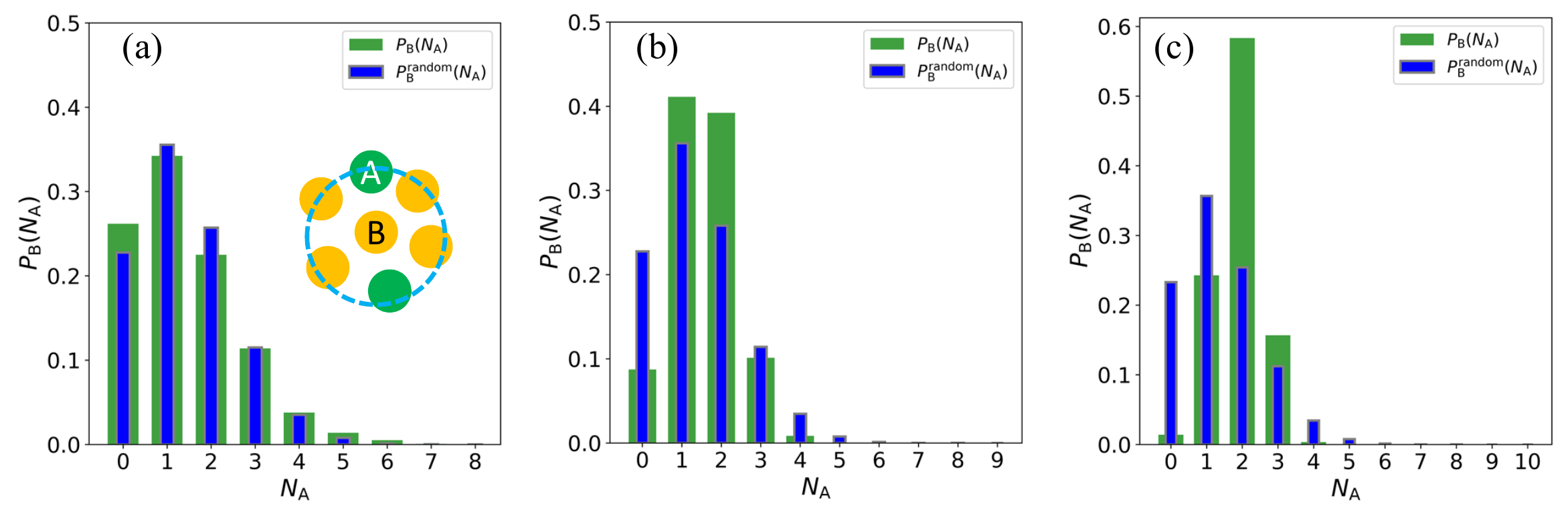}
\caption{The probability $P_B(N_A)$ for a $B$ atom to have $N_A$ (Voronoi) 
nearest neighbors for $f_B = 0.9$ in binary LJ systems cooled at 
$R=5 \times 10^{-5}$ (green wide 
bars) for combinations 
of $\epsilon_-$ and ${\overline \epsilon}_{AB}$ that yield (a) $R_c = 6.6 \times
10^{-4}$ (${\overline \epsilon}_{AB}=0.83$, $\epsilon_-=-0.67$), 
(b) $5.6 \times 10^{-5}$ (${\overline \epsilon}_{AB}=4.17$, $\epsilon_-=-0.11$), and 
(c) $<10^{-6}$ (${\overline \epsilon}_{AB}=5.0$, $\epsilon_-=-0.67$). We also show 
$P_B^{\rm random}(N_A)$ (thin blue bars) for systems with the same structure, but randomized 
atomic labels for the nearest neighbors.}
\label{fig5}
\end{figure*}

In the high-temperature liquid, the identities of the nearest
(Voronoi) neighbors of atoms $A$ and $B$ are completely random. As the
system cools, the identities of the neighboring atoms for each atom
type $A$ and $B$ can deviate from random, and such chemical ordering
can affect the GFA. For example, we hypothesize that if the competing
crystal has large chemical order, the system will possess large GFA
since the $A$ and $B$ species must rearrange significantly to form the
crystal. To assess this hypothesis, we measured the chemical ordering
(i.e. the probability $p_A(N_B)$ for an $A$ atom to have $N_B$ $B$
nearest neighbors when $f_A > f_B$ or the probability $p_B(N_A)$ for a
$B$ atom to have $N_A$ $A$ nearest neighbors when $f_B > f_A$) 
at a slow cooling rate with
signficant FCC order. In Fig.~\ref{fig5}, we show $p_B(N_A)$ for
systems with $f_B = 0.9$ and $R_c$ decreasing from (a) to (c). We
compare $p_B(N_A)$ to $p_B^{\rm random}(N_A)$, where we keep the
low-temperature structure of the system and randomly assign the labels
of the nearest neighbors.  We find that the GFA increases with
the chemical order, $\sum_{N_A} |p_B(N_A)-p_{B}^{\rm random}(N_A)|$,
of the competing crystal. We find similar results for
systems with $f_B = 0.1$; the GFA increases with the chemical order,
$\sum_{N_B} |p_A(N_B)-p_{A}^{\rm random}(N_B)|$, of the competing
crystal. (See Supplemental Material.)

By decoupling geometric and energetic frustration, we have shown that
the GFA is not strongly correlated to the heat of mixing, which
involves the particular combination of variables, $(\epsilon_{BB}
+\epsilon_{AA})/2 \epsilon_{AA}
-\epsilon_{AB}/\epsilon_{AA}$. Instead, we find that the GFA is
strongly correlated with $\epsilon_-$ (i.e. the difference in the
cohesive energies, not the sum) and ${\overline \epsilon}_{AB}$, and
we identified the $f_B$-dependent combination of $\epsilon_-$ and
${\overline \epsilon}_{AB}$ that controls the GFA for binary LJ
systems. We emphasize that it was important to study regions of the
$\epsilon_-$ and ${\overline \epsilon}_{AB}$ parameter space that were
beyond the experimental range of metallic glasses to fully understand
the GFA.  This work will motivate several important future
studies. First, we encourage researchers to experimentally
characterize the GFA of binary alloys containing nearly monoatomic
elements, yet with large energetic frustration. Second, we are now in
a position to understand theoretically the GFA of binary LJ systems
with both geometric and energetc frustration. For example, it will be
interesting to determine how energetic frustration couples to
geometric frustration. For example, should element $A$ with larger
cohesive energy possess a larger or smaller metallic radius than element
$B$ to yield large GFA?

\section*{Acknowledgements}

The authors acknowledge support from NSF MRSEC Grant No. DMR-1119826
(Y.-C.H.) and NSF Grant Nos. CMMI-1462439 (C.O.) and CMMI-1463455
(M.S.). This work was supported by the High Performance Computing
facilities operated by, and the staff of, the Yale Center for Research
Computing.



\balance


\bibliography{rsc_sa} 

\begin{thebibliography}{31}%
\makeatletter
\providecommand \@ifxundefined [1]{%
 \@ifx{#1\undefined}
}%
\providecommand \@ifnum [1]{%
 \ifnum #1\expandafter \@firstoftwo
 \else \expandafter \@secondoftwo
 \fi
}%
\providecommand \@ifx [1]{%
 \ifx #1\expandafter \@firstoftwo
 \else \expandafter \@secondoftwo
 \fi
}%
\providecommand \natexlab [1]{#1}%
\providecommand \enquote  [1]{``#1''}%
\providecommand \bibnamefont  [1]{#1}%
\providecommand \bibfnamefont [1]{#1}%
\providecommand \citenamefont [1]{#1}%
\providecommand \href@noop [0]{\@secondoftwo}%
\providecommand \href [0]{\begingroup \@sanitize@url \@href}%
\providecommand \@href[1]{\@@startlink{#1}\@@href}%
\providecommand \@@href[1]{\endgroup#1\@@endlink}%
\providecommand \@sanitize@url [0]{\catcode `\\12\catcode `\$12\catcode
  `\&12\catcode `\#12\catcode `\^12\catcode `\_12\catcode `\%12\relax}%
\providecommand \@@startlink[1]{}%
\providecommand \@@endlink[0]{}%
\providecommand \url  [0]{\begingroup\@sanitize@url \@url }%
\providecommand \@url [1]{\endgroup\@href {#1}{\urlprefix }}%
\providecommand \urlprefix  [0]{URL }%
\providecommand \Eprint [0]{\href }%
\providecommand \doibase [0]{http://dx.doi.org/}%
\providecommand \selectlanguage [0]{\@gobble}%
\providecommand \bibinfo  [0]{\@secondoftwo}%
\providecommand \bibfield  [0]{\@secondoftwo}%
\providecommand \translation [1]{[#1]}%
\providecommand \BibitemOpen [0]{}%
\providecommand \bibitemStop [0]{}%
\providecommand \bibitemNoStop [0]{.\EOS\space}%
\providecommand \EOS [0]{\spacefactor3000\relax}%
\providecommand \BibitemShut  [1]{\csname bibitem#1\endcsname}%
\let\auto@bib@innerbib\@empty
\bibitem [{\citenamefont {Schroers}(2013)}]{JanPT}%
  \BibitemOpen
  \bibfield  {author} {\bibinfo {author} {\bibfnamefont {Jan}\ \bibnamefont
  {Schroers}},\ }\bibfield  {title} {\enquote {\bibinfo {title} {Bulk metallic
  glasses},}\ }\href {\doibase 10.1063/PT.3.1885} {\bibfield  {journal}
  {\bibinfo  {journal} {Phys. Today}\ }\textbf {\bibinfo {volume} {66}},\
  \bibinfo {pages} {32} (\bibinfo {year} {2013})}\BibitemShut {NoStop}%
\bibitem [{\citenamefont {Demetriou}\ \emph {et~al.}(2011)\citenamefont
  {Demetriou}, \citenamefont {Launey}, \citenamefont {Garrett}, \citenamefont
  {Schramm}, \citenamefont {Hofmann}, \citenamefont {Johnson},\ and\
  \citenamefont {Ritchie}}]{Demetriou2011}%
  \BibitemOpen
  \bibfield  {author} {\bibinfo {author} {\bibfnamefont {Marios~D.}\
  \bibnamefont {Demetriou}}, \bibinfo {author} {\bibfnamefont {Maximilien~E.}\
  \bibnamefont {Launey}}, \bibinfo {author} {\bibfnamefont {Glenn}\
  \bibnamefont {Garrett}}, \bibinfo {author} {\bibfnamefont {Joseph~P.}\
  \bibnamefont {Schramm}}, \bibinfo {author} {\bibfnamefont {Douglas~C.}\
  \bibnamefont {Hofmann}}, \bibinfo {author} {\bibfnamefont {William~L.}\
  \bibnamefont {Johnson}}, \ and\ \bibinfo {author} {\bibfnamefont {Robert~O.}\
  \bibnamefont {Ritchie}},\ }\bibfield  {title} {\enquote {\bibinfo {title} {A
  damage-tolerant glass},}\ }\href {https://doi.org/10.1038/nmat2930}
  {\bibfield  {journal} {\bibinfo  {journal} {Nat. Mater.}\ }\textbf {\bibinfo
  {volume} {10}},\ \bibinfo {pages} {123} (\bibinfo {year} {2011})}\BibitemShut
  {NoStop}%
\bibitem [{\citenamefont {Wang}(2012)}]{WANG2012487}%
  \BibitemOpen
  \bibfield  {author} {\bibinfo {author} {\bibfnamefont {Wei~Hua}\ \bibnamefont
  {Wang}},\ }\bibfield  {title} {\enquote {\bibinfo {title} {The elastic
  properties, elastic models and elastic perspectives of metallic glasses},}\
  }\href {\doibase https://doi.org/10.1016/j.pmatsci.2011.07.001} {\bibfield
  {journal} {\bibinfo  {journal} {Prog. Mater. Sci.}\ }\textbf {\bibinfo
  {volume} {57}},\ \bibinfo {pages} {487} (\bibinfo {year} {2012})}\BibitemShut
  {NoStop}%
\bibitem [{\citenamefont {Zberg}\ \emph {et~al.}(2009)\citenamefont {Zberg},
  \citenamefont {Uggowitzer},\ and\ \citenamefont {L{\"o}ffler}}]{Zberg2009}%
  \BibitemOpen
  \bibfield  {author} {\bibinfo {author} {\bibfnamefont {Bruno}\ \bibnamefont
  {Zberg}}, \bibinfo {author} {\bibfnamefont {Peter~J.}\ \bibnamefont
  {Uggowitzer}}, \ and\ \bibinfo {author} {\bibfnamefont {J{\"o}rg~F.}\
  \bibnamefont {L{\"o}ffler}},\ }\bibfield  {title} {\enquote {\bibinfo {title}
  {Mgznca glasses without clinically observable hydrogen evolution for
  biodegradable implants},}\ }\href {https://doi.org/10.1038/nmat2542}
  {\bibfield  {journal} {\bibinfo  {journal} {Nat. Mater.}\ }\textbf {\bibinfo
  {volume} {8}},\ \bibinfo {pages} {887} (\bibinfo {year} {2009})}\BibitemShut
  {NoStop}%
\bibitem [{\citenamefont {Lu}\ and\ \citenamefont {Liu}(2002)}]{LU20023501}%
  \BibitemOpen
  \bibfield  {author} {\bibinfo {author} {\bibfnamefont {Z.P.}\ \bibnamefont
  {Lu}}\ and\ \bibinfo {author} {\bibfnamefont {C.T.}\ \bibnamefont {Liu}},\
  }\bibfield  {title} {\enquote {\bibinfo {title} {A new glass-forming ability
  criterion for bulk metallic glasses},}\ }\href@noop {} {\bibfield  {journal}
  {\bibinfo  {journal} {Acta Mater.}\ }\textbf {\bibinfo {volume} {50}},\
  \bibinfo {pages} {3501} (\bibinfo {year} {2002})}\BibitemShut {NoStop}%
\bibitem [{\citenamefont {Inoue}(2000)}]{INOUE2000279}%
  \BibitemOpen
  \bibfield  {author} {\bibinfo {author} {\bibfnamefont {Akihisa}\ \bibnamefont
  {Inoue}},\ }\bibfield  {title} {\enquote {\bibinfo {title} {Stabilization of
  metallic supercooled liquid and bulk amorphous alloys},}\ }\href {\doibase
  https://doi.org/10.1016/S1359-6454(99)00300-6} {\bibfield  {journal}
  {\bibinfo  {journal} {Acta Mater.}\ }\textbf {\bibinfo {volume} {48}},\
  \bibinfo {pages} {279} (\bibinfo {year} {2000})}\BibitemShut {NoStop}%
\bibitem [{\citenamefont {Xu}\ \emph {et~al.}(2004)\citenamefont {Xu},
  \citenamefont {Lohwongwatana}, \citenamefont {Duan}, \citenamefont
  {Johnson},\ and\ \citenamefont {Garland}}]{XU20042621}%
  \BibitemOpen
  \bibfield  {author} {\bibinfo {author} {\bibfnamefont {Donghua}\ \bibnamefont
  {Xu}}, \bibinfo {author} {\bibfnamefont {Boonrat}\ \bibnamefont
  {Lohwongwatana}}, \bibinfo {author} {\bibfnamefont {Gang}\ \bibnamefont
  {Duan}}, \bibinfo {author} {\bibfnamefont {William~L.}\ \bibnamefont
  {Johnson}}, \ and\ \bibinfo {author} {\bibfnamefont {Carol}\ \bibnamefont
  {Garland}},\ }\bibfield  {title} {\enquote {\bibinfo {title} {Bulk metallic
  glass formation in binary cu-rich alloy series - {C}u$_{100-x}${Z}r$_{x}$
  (x=34, 36, 38.2, 40 at.\%) and mechanical properties of bulk
  {C}u$_{64}${Z}r$_{36}$ glass},}\ }\href {\doibase
  https://doi.org/10.1016/j.actamat.2004.02.009} {\bibfield  {journal}
  {\bibinfo  {journal} {Acta Mater.}\ }\textbf {\bibinfo {volume} {52}},\
  \bibinfo {pages} {2621} (\bibinfo {year} {2004})}\BibitemShut {NoStop}%
\bibitem [{\citenamefont {Li}\ \emph {et~al.}(2008)\citenamefont {Li},
  \citenamefont {Guo}, \citenamefont {Kalb},\ and\ \citenamefont
  {Thompson}}]{Li1816}%
  \BibitemOpen
  \bibfield  {author} {\bibinfo {author} {\bibfnamefont {Y.}~\bibnamefont
  {Li}}, \bibinfo {author} {\bibfnamefont {Q.}~\bibnamefont {Guo}}, \bibinfo
  {author} {\bibfnamefont {J.~A.}\ \bibnamefont {Kalb}}, \ and\ \bibinfo
  {author} {\bibfnamefont {C.~V.}\ \bibnamefont {Thompson}},\ }\bibfield
  {title} {\enquote {\bibinfo {title} {Matching glass-forming ability with the
  density of the amorphous phase},}\ }\href {\doibase 10.1126/science.1163062}
  {\bibfield  {journal} {\bibinfo  {journal} {Science}\ }\textbf {\bibinfo
  {volume} {322}},\ \bibinfo {pages} {1816} (\bibinfo {year}
  {2008})}\BibitemShut {NoStop}%
\bibitem [{\citenamefont {Tang}\ \emph {et~al.}(2004)\citenamefont {Tang},
  \citenamefont {Zhao}, \citenamefont {Pan},\ and\ \citenamefont
  {Wang}}]{mei2004binary}%
  \BibitemOpen
  \bibfield  {author} {\bibinfo {author} {\bibfnamefont {M.~B.}\ \bibnamefont
  {Tang}}, \bibinfo {author} {\bibfnamefont {D.~Q.}\ \bibnamefont {Zhao}},
  \bibinfo {author} {\bibfnamefont {M.~X.}\ \bibnamefont {Pan}}, \ and\
  \bibinfo {author} {\bibfnamefont {W.~H.}\ \bibnamefont {Wang}},\ }\bibfield
  {title} {\enquote {\bibinfo {title} {Binary {C}u-{Z}r bulk metallic
  glasses},}\ }\href {\doibase 10.1088/0256-307x/21/5/039} {\bibfield
  {journal} {\bibinfo  {journal} {Chi. Phys. Lett.}\ }\textbf {\bibinfo
  {volume} {21}},\ \bibinfo {pages} {901} (\bibinfo {year} {2004})}\BibitemShut
  {NoStop}%
\bibitem [{\citenamefont {Tsai}\ \emph {et~al.}(1988)\citenamefont {Tsai},
  \citenamefont {Inoue},\ and\ \citenamefont {Masumoto}}]{tsai1988ductile}%
  \BibitemOpen
  \bibfield  {author} {\bibinfo {author} {\bibfnamefont {An-Pang}\ \bibnamefont
  {Tsai}}, \bibinfo {author} {\bibfnamefont {Akihisa}\ \bibnamefont {Inoue}}, \
  and\ \bibinfo {author} {\bibfnamefont {Tsuyoshi}\ \bibnamefont {Masumoto}},\
  }\bibfield  {title} {\enquote {\bibinfo {title} {Ductile al-cu-v amorphous
  alloys without metalloid},}\ }\href@noop {} {\bibfield  {journal} {\bibinfo
  {journal} {Metall. Trans. A}\ }\textbf {\bibinfo {volume} {19}},\ \bibinfo
  {pages} {391} (\bibinfo {year} {1988})}\BibitemShut {NoStop}%
\bibitem [{\citenamefont {Zhong}\ \emph {et~al.}(2014)\citenamefont {Zhong},
  \citenamefont {Wang}, \citenamefont {Sheng}, \citenamefont {Zhang},\ and\
  \citenamefont {Mao}}]{Zhong2014}%
  \BibitemOpen
  \bibfield  {author} {\bibinfo {author} {\bibfnamefont {Li}~\bibnamefont
  {Zhong}}, \bibinfo {author} {\bibfnamefont {Jiangwei}\ \bibnamefont {Wang}},
  \bibinfo {author} {\bibfnamefont {Hongwei}\ \bibnamefont {Sheng}}, \bibinfo
  {author} {\bibfnamefont {Ze}~\bibnamefont {Zhang}}, \ and\ \bibinfo {author}
  {\bibfnamefont {Scott~X.}\ \bibnamefont {Mao}},\ }\bibfield  {title}
  {\enquote {\bibinfo {title} {Formation of monatomic metallic glasses through
  ultrafast liquid quenching},}\ }\href {https://doi.org/10.1038/nature13617}
  {\bibfield  {journal} {\bibinfo  {journal} {Nature}\ }\textbf {\bibinfo
  {volume} {512}},\ \bibinfo {pages} {177} (\bibinfo {year}
  {2014})}\BibitemShut {NoStop}%
\bibitem [{\citenamefont {Zhang}\ \emph {et~al.}(2013)\citenamefont {Zhang},
  \citenamefont {Wang}, \citenamefont {Papanikolaou}, \citenamefont {Liu},
  \citenamefont {Schroers}, \citenamefont {Shattuck},\ and\ \citenamefont
  {O'Hern}}]{KZ2013}%
  \BibitemOpen
  \bibfield  {author} {\bibinfo {author} {\bibfnamefont {Kai}\ \bibnamefont
  {Zhang}}, \bibinfo {author} {\bibfnamefont {Minglei}\ \bibnamefont {Wang}},
  \bibinfo {author} {\bibfnamefont {Stefanos}\ \bibnamefont {Papanikolaou}},
  \bibinfo {author} {\bibfnamefont {Yanhui}\ \bibnamefont {Liu}}, \bibinfo
  {author} {\bibfnamefont {Jan}\ \bibnamefont {Schroers}}, \bibinfo {author}
  {\bibfnamefont {Mark~D.}\ \bibnamefont {Shattuck}}, \ and\ \bibinfo {author}
  {\bibfnamefont {Corey~S.}\ \bibnamefont {O'Hern}},\ }\bibfield  {title}
  {\enquote {\bibinfo {title} {Computational studies of the glass-forming
  ability of model bulk metallic glasses},}\ }\href {\doibase
  10.1063/1.4821637} {\bibfield  {journal} {\bibinfo  {journal} {J. Chem.
  Phys.}\ }\textbf {\bibinfo {volume} {139}},\ \bibinfo {pages} {124503}
  (\bibinfo {year} {2013})}\BibitemShut {NoStop}%
\bibitem [{\citenamefont {Shintani}\ and\ \citenamefont
  {Tanaka}(2006)}]{Tanaka2006NP}%
  \BibitemOpen
  \bibfield  {author} {\bibinfo {author} {\bibfnamefont {Hiroshi}\ \bibnamefont
  {Shintani}}\ and\ \bibinfo {author} {\bibfnamefont {Hajime}\ \bibnamefont
  {Tanaka}},\ }\bibfield  {title} {\enquote {\bibinfo {title} {Frustration on
  the way to crystallization in glass},}\ }\href@noop {} {\bibfield  {journal}
  {\bibinfo  {journal} {Nat. Phys.}\ }\textbf {\bibinfo {volume} {2}},\
  \bibinfo {pages} {200} (\bibinfo {year} {2006})}\BibitemShut {NoStop}%
\bibitem [{\citenamefont {Miracle}(2004)}]{Miracle2004}%
  \BibitemOpen
  \bibfield  {author} {\bibinfo {author} {\bibfnamefont {Daniel~B.}\
  \bibnamefont {Miracle}},\ }\bibfield  {title} {\enquote {\bibinfo {title} {A
  structural model for metallic glasses},}\ }\href {\doibase 10.1038/nmat1219}
  {\bibfield  {journal} {\bibinfo  {journal} {Nat. Mater.}\ }\textbf {\bibinfo
  {volume} {3}},\ \bibinfo {pages} {697} (\bibinfo {year} {2004})}\BibitemShut
  {NoStop}%
\bibitem [{\citenamefont {Sheng}\ \emph {et~al.}(2006)\citenamefont {Sheng},
  \citenamefont {Luo}, \citenamefont {Alamgir}, \citenamefont {Bai},\ and\
  \citenamefont {Ma}}]{Sheng2006}%
  \BibitemOpen
  \bibfield  {author} {\bibinfo {author} {\bibfnamefont {H.~W.}\ \bibnamefont
  {Sheng}}, \bibinfo {author} {\bibfnamefont {W.~K.}\ \bibnamefont {Luo}},
  \bibinfo {author} {\bibfnamefont {F.~M.}\ \bibnamefont {Alamgir}}, \bibinfo
  {author} {\bibfnamefont {J.~M.}\ \bibnamefont {Bai}}, \ and\ \bibinfo
  {author} {\bibfnamefont {E.}~\bibnamefont {Ma}},\ }\bibfield  {title}
  {\enquote {\bibinfo {title} {Atomic packing and short-to-medium-range order
  in metallic glasses},}\ }\href {\doibase 10.1038/nature04421} {\bibfield
  {journal} {\bibinfo  {journal} {Nature}\ }\textbf {\bibinfo {volume} {439}},\
  \bibinfo {pages} {419} (\bibinfo {year} {2006})}\BibitemShut {NoStop}%
\bibitem [{\citenamefont {Hu}\ \emph {et~al.}(2015)\citenamefont {Hu},
  \citenamefont {Li}, \citenamefont {Li}, \citenamefont {Bai},\ and\
  \citenamefont {Wang}}]{HU2015NC}%
  \BibitemOpen
  \bibfield  {author} {\bibinfo {author} {\bibfnamefont {Y.~C.}\ \bibnamefont
  {Hu}}, \bibinfo {author} {\bibfnamefont {F.~X.}\ \bibnamefont {Li}}, \bibinfo
  {author} {\bibfnamefont {M.~Z.}\ \bibnamefont {Li}}, \bibinfo {author}
  {\bibfnamefont {H.~Y.}\ \bibnamefont {Bai}}, \ and\ \bibinfo {author}
  {\bibfnamefont {W.~H.}\ \bibnamefont {Wang}},\ }\bibfield  {title} {\enquote
  {\bibinfo {title} {Five-fold symmetry as indicator of dynamic arrest in
  metallic glass-forming liquids},}\ }\href {\doibase
  https://doi.org/10.1038/ncomms9310} {\bibfield  {journal} {\bibinfo
  {journal} {Nat. Commun.}\ }\textbf {\bibinfo {volume} {6}},\ \bibinfo {pages}
  {8310} (\bibinfo {year} {2015})}\BibitemShut {NoStop}%
\bibitem [{\citenamefont {Cheng}\ \emph {et~al.}(2009)\citenamefont {Cheng},
  \citenamefont {Ma},\ and\ \citenamefont {Sheng}}]{ChengPRL2009}%
  \BibitemOpen
  \bibfield  {author} {\bibinfo {author} {\bibfnamefont {Y.~Q.}\ \bibnamefont
  {Cheng}}, \bibinfo {author} {\bibfnamefont {E.}~\bibnamefont {Ma}}, \ and\
  \bibinfo {author} {\bibfnamefont {H.~W.}\ \bibnamefont {Sheng}},\ }\bibfield
  {title} {\enquote {\bibinfo {title} {Atomic level structure in multicomponent
  bulk metallic glass},}\ }\href {\doibase 10.1103/PhysRevLett.102.245501}
  {\bibfield  {journal} {\bibinfo  {journal} {Phys. Rev. Lett.}\ }\textbf
  {\bibinfo {volume} {102}},\ \bibinfo {pages} {245501} (\bibinfo {year}
  {2009})}\BibitemShut {NoStop}%
\bibitem [{\citenamefont {Cargill}\ and\ \citenamefont
  {Spaepen}(1981)}]{chemicalorder}%
  \BibitemOpen
  \bibfield  {author} {\bibinfo {author} {\bibfnamefont {G.S.}\ \bibnamefont
  {Cargill}}\ and\ \bibinfo {author} {\bibfnamefont {F.}~\bibnamefont
  {Spaepen}},\ }\bibfield  {title} {\enquote {\bibinfo {title} {Description of
  chemical ordering in amorphous alloys},}\ }\href {\doibase
  https://doi.org/10.1016/0022-3093(81)90174-5} {\bibfield  {journal} {\bibinfo
   {journal} {J. Non-Cryst. Solids}\ }\textbf {\bibinfo {volume} {43}},\
  \bibinfo {pages} {91} (\bibinfo {year} {1981})}\BibitemShut {NoStop}%
\bibitem [{\citenamefont {Nos\'{e}}(1984)}]{nose}%
  \BibitemOpen
  \bibfield  {author} {\bibinfo {author} {\bibfnamefont {Shuichi}\ \bibnamefont
  {Nos\'{e}}},\ }\bibfield  {title} {\enquote {\bibinfo {title} {A unified
  formulation of the constant temperature molecular dynamics methods},}\ }\href
  {\doibase 10.1063/1.447334} {\bibfield  {journal} {\bibinfo  {journal} {J.
  Chem. Phys.}\ }\textbf {\bibinfo {volume} {81}},\ \bibinfo {pages} {511}
  (\bibinfo {year} {1984})}\BibitemShut {NoStop}%
\bibitem [{\citenamefont {Hoover}(1985)}]{hoover}%
  \BibitemOpen
  \bibfield  {author} {\bibinfo {author} {\bibfnamefont {William~G.}\
  \bibnamefont {Hoover}},\ }\bibfield  {title} {\enquote {\bibinfo {title}
  {Canonical dynamics: equilibrium phase-space distributions},}\ }\href
  {\doibase 10.1103/PhysRevA.31.1695} {\bibfield  {journal} {\bibinfo
  {journal} {Phys. Rev. A}\ }\textbf {\bibinfo {volume} {31}},\ \bibinfo
  {pages} {1695} (\bibinfo {year} {1985})}\BibitemShut {NoStop}%
\bibitem [{\citenamefont {Martyna}\ \emph {et~al.}(1994)\citenamefont
  {Martyna}, \citenamefont {Tobias},\ and\ \citenamefont {Klein}}]{npt}%
  \BibitemOpen
  \bibfield  {author} {\bibinfo {author} {\bibfnamefont {Glenn~J.}\
  \bibnamefont {Martyna}}, \bibinfo {author} {\bibfnamefont {Douglas~J.}\
  \bibnamefont {Tobias}}, \ and\ \bibinfo {author} {\bibfnamefont {Michael~L.}\
  \bibnamefont {Klein}},\ }\bibfield  {title} {\enquote {\bibinfo {title}
  {Constant pressure molecular dynamics algorithms},}\ }\href {\doibase
  10.1063/1.467468} {\bibfield  {journal} {\bibinfo  {journal} {J. Chem.
  Phys.}\ }\textbf {\bibinfo {volume} {101}},\ \bibinfo {pages} {4177}
  (\bibinfo {year} {1994})}\BibitemShut {NoStop}%
\bibitem [{\citenamefont {Halpern}(2012)}]{cohesive}%
  \BibitemOpen
  \bibfield  {author} {\bibinfo {author} {\bibfnamefont {Arthur~M.}\
  \bibnamefont {Halpern}},\ }\bibfield  {title} {\enquote {\bibinfo {title}
  {From dimer to crystal: calculating the cohesive energy of rare gas
  solids},}\ }\href {\doibase 10.1021/ed200348j} {\bibfield  {journal}
  {\bibinfo  {journal} {J. Chem. Educ.}\ }\textbf {\bibinfo {volume} {89}},\
  \bibinfo {pages} {592} (\bibinfo {year} {2012})}\BibitemShut {NoStop}%
\bibitem [{\citenamefont {Takeuchi}\ and\ \citenamefont
  {Inoue}(2005)}]{AA2005}%
  \BibitemOpen
  \bibfield  {author} {\bibinfo {author} {\bibfnamefont {Akira}\ \bibnamefont
  {Takeuchi}}\ and\ \bibinfo {author} {\bibfnamefont {Akihisa}\ \bibnamefont
  {Inoue}},\ }\bibfield  {title} {\enquote {\bibinfo {title} {Classification of
  bulk metallic glasses by atomic size difference, heat of mixing and period of
  constituent elements and its application to characterization of the main
  alloying element},}\ }\href {\doibase 10.2320/matertrans.46.2817} {\bibfield
  {journal} {\bibinfo  {journal} {Mater. Trans.}\ }\textbf {\bibinfo {volume}
  {46}},\ \bibinfo {pages} {2817} (\bibinfo {year} {2005})}\BibitemShut
  {NoStop}%
\bibitem [{\citenamefont {Takeuchi}\ and\ \citenamefont {Inoue}(2000)}]{AAMix}%
  \BibitemOpen
  \bibfield  {author} {\bibinfo {author} {\bibfnamefont {Akira}\ \bibnamefont
  {Takeuchi}}\ and\ \bibinfo {author} {\bibfnamefont {Akihisa}\ \bibnamefont
  {Inoue}},\ }\bibfield  {title} {\enquote {\bibinfo {title} {Calculations of
  mixing enthalpy and mismatch entropy for ternary amorphous alloys},}\ }\href
  {\doibase 10.2320/matertrans1989.41.1372} {\bibfield  {journal} {\bibinfo
  {journal} {Mater. Trans.}\ }\textbf {\bibinfo {volume} {41}},\ \bibinfo
  {pages} {1372} (\bibinfo {year} {2000})}\BibitemShut {NoStop}%
\bibitem [{\citenamefont {Köddermann}\ \emph {et~al.}(2007)\citenamefont
  {Köddermann}, \citenamefont {Paschek},\ and\ \citenamefont
  {Ludwig}}]{ionicliquid}%
  \BibitemOpen
  \bibfield  {author} {\bibinfo {author} {\bibfnamefont {Thorsten}\
  \bibnamefont {Köddermann}}, \bibinfo {author} {\bibfnamefont {Dietmar}\
  \bibnamefont {Paschek}}, \ and\ \bibinfo {author} {\bibfnamefont {Ralf}\
  \bibnamefont {Ludwig}},\ }\bibfield  {title} {\enquote {\bibinfo {title}
  {Molecular dynamic simulations of ionic liquids: a reliable description of
  structure, thermodynamics and dynamics},}\ }\href {\doibase
  10.1002/cphc.200700552} {\bibfield  {journal} {\bibinfo  {journal}
  {ChemPhysChem}\ }\textbf {\bibinfo {volume} {8}},\ \bibinfo {pages} {2464}
  (\bibinfo {year} {2007})}\BibitemShut {NoStop}%
\bibitem [{\citenamefont {de~Andrade}\ \emph {et~al.}(2002)\citenamefont
  {de~Andrade}, \citenamefont {Böes},\ and\ \citenamefont {Stassen}}]{molten}%
  \BibitemOpen
  \bibfield  {author} {\bibinfo {author} {\bibfnamefont {Jones}\ \bibnamefont
  {de~Andrade}}, \bibinfo {author} {\bibfnamefont {Elvis~S.}\ \bibnamefont
  {Böes}}, \ and\ \bibinfo {author} {\bibfnamefont {Hubert}\ \bibnamefont
  {Stassen}},\ }\bibfield  {title} {\enquote {\bibinfo {title} {Computational
  study of room temperature molten salts composed by
  1-alkyl-3-methylimidazolium cations- force-field proposal and validation},}\
  }\href {\doibase 10.1021/jp0216629} {\bibfield  {journal} {\bibinfo
  {journal} {J. Phys. Chem. B}\ }\textbf {\bibinfo {volume} {106}},\ \bibinfo
  {pages} {13344} (\bibinfo {year} {2002})}\BibitemShut {NoStop}%
\bibitem [{\citenamefont {London}(1937)}]{london}%
  \BibitemOpen
  \bibfield  {author} {\bibinfo {author} {\bibfnamefont {F.}~\bibnamefont
  {London}},\ }\bibfield  {title} {\enquote {\bibinfo {title} {The general
  theory of molecular forces},}\ }\href {\doibase 10.1039/TF937330008B}
  {\bibfield  {journal} {\bibinfo  {journal} {Trans. Faraday Soc.}\ }\textbf
  {\bibinfo {volume} {33}},\ \bibinfo {pages} {8} (\bibinfo {year}
  {1937})}\BibitemShut {NoStop}%
\bibitem [{\citenamefont {Steinhardt}\ \emph {et~al.}(1983)\citenamefont
  {Steinhardt}, \citenamefont {Nelson},\ and\ \citenamefont {Ronchetti}}]{BOO}%
  \BibitemOpen
  \bibfield  {author} {\bibinfo {author} {\bibfnamefont {Paul~J.}\ \bibnamefont
  {Steinhardt}}, \bibinfo {author} {\bibfnamefont {David~R.}\ \bibnamefont
  {Nelson}}, \ and\ \bibinfo {author} {\bibfnamefont {Marco}\ \bibnamefont
  {Ronchetti}},\ }\bibfield  {title} {\enquote {\bibinfo {title}
  {Bond-orientational order in liquids and glasses},}\ }\href {\doibase
  10.1103/PhysRevB.28.784} {\bibfield  {journal} {\bibinfo  {journal} {Phys.
  Rev. B}\ }\textbf {\bibinfo {volume} {28}},\ \bibinfo {pages} {784} (\bibinfo
  {year} {1983})}\BibitemShut {NoStop}%
\bibitem [{\citenamefont {Mickel}\ \emph {et~al.}(2013)\citenamefont {Mickel},
  \citenamefont {Kapfer}, \citenamefont {Schr\"{o}der-Turk},\ and\
  \citenamefont {Mecke}}]{BOOVoronoi}%
  \BibitemOpen
  \bibfield  {author} {\bibinfo {author} {\bibfnamefont {Walter}\ \bibnamefont
  {Mickel}}, \bibinfo {author} {\bibfnamefont {Sebastian~C.}\ \bibnamefont
  {Kapfer}}, \bibinfo {author} {\bibfnamefont {Gerd~E.}\ \bibnamefont
  {Schr\"{o}der-Turk}}, \ and\ \bibinfo {author} {\bibfnamefont {Klaus}\
  \bibnamefont {Mecke}},\ }\bibfield  {title} {\enquote {\bibinfo {title}
  {Shortcomings of the bond orientational order parameters for the analysis of
  disordered particulate matter},}\ }\href {\doibase 10.1063/1.4774084}
  {\bibfield  {journal} {\bibinfo  {journal} {J. Chem. Phys.}\ }\textbf
  {\bibinfo {volume} {138}},\ \bibinfo {pages} {044501} (\bibinfo {year}
  {2013})}\BibitemShut {NoStop}%
\bibitem [{\citenamefont {Stukowski}(2012)}]{cna}%
  \BibitemOpen
  \bibfield  {author} {\bibinfo {author} {\bibfnamefont {Alexander}\
  \bibnamefont {Stukowski}},\ }\bibfield  {title} {\enquote {\bibinfo {title}
  {Structure identification methods for atomistic simulations of crystalline
  materials},}\ }\href {\doibase 10.1088/0965-0393/20/4/045021} {\bibfield
  {journal} {\bibinfo  {journal} {Model. Simul. Mater. Sci. Eng}\ }\textbf
  {\bibinfo {volume} {20}},\ \bibinfo {pages} {045021} (\bibinfo {year}
  {2012})}\BibitemShut {NoStop}%
\bibitem [{\citenamefont {Li}\ \emph {et~al.}(2017)\citenamefont {Li},
  \citenamefont {Zhao}, \citenamefont {Liu}, \citenamefont {Gong},\ and\
  \citenamefont {Schroers}}]{liH}%
  \BibitemOpen
  \bibfield  {author} {\bibinfo {author} {\bibfnamefont {Yanglin}\ \bibnamefont
  {Li}}, \bibinfo {author} {\bibfnamefont {Shaofan}\ \bibnamefont {Zhao}},
  \bibinfo {author} {\bibfnamefont {Yanhui}\ \bibnamefont {Liu}}, \bibinfo
  {author} {\bibfnamefont {Pan}\ \bibnamefont {Gong}}, \ and\ \bibinfo {author}
  {\bibfnamefont {Jan}\ \bibnamefont {Schroers}},\ }\bibfield  {title}
  {\enquote {\bibinfo {title} {How many bulk metallic glasses are there?}}\
  }\href {\doibase 10.1021/acscombsci.7b00048} {\bibfield  {journal} {\bibinfo
  {journal} {ACS Comb. Sci.}\ }\textbf {\bibinfo {volume} {19}},\ \bibinfo
  {pages} {687} (\bibinfo {year} {2017})}\BibitemShut {NoStop}%
\end{thebibliography}%

\end{document}